\documentclass[doublecol]{epl2}
\usepackage{amsmath,amssymb,hyperref}

\newcommand{\Tr}{{\rm Tr}}

\newcommand{\be}{\begin{equation}}
\newcommand{\ee}{\end{equation}}

\begin{document}

\title{Time delay statistics for finite number of channels in all symmetry classes}
\author{Marcel Novaes}
\institute{Instituto de F\'isica, Universidade Federal de Uberl\^andia, 38408-100, Brazil}
\abstract{
Within a random matrix theory approach, we obtain spectral statistics of the Wigner time delay matrix $Q$, for arbitrary channels number $M$ and for all symmetry classes, in fact for general Dyson parameter $\beta$. We also put forth two conjectures: one is related to the large-$M$ expansion of joint cumulants of traces of powers of $Q$, which generalizes and implies a previous conjecture of Cunden, Mezzadri, Vivo and Simm; the other concerns the tail of the distribution of traces of powers of $Q$.
}

\maketitle

\section{Introduction} 

Time has always been a difficult concept in quantum physics \cite{wigner,smith,review1,review2}. One attempt to quantify the duration of a scattering process is to introduce the Wigner-Smith time delay matrix $Q=-i\hbar S^\dagger \frac{dS}{dE}$, where $S$ is the scattering matrix and $E$ is the energy. 

Then the classical dynamics is chaotic, the matrix elements of $Q$ are widely fluctating functions of energy. Therefore, it makes sense to introduce a local energy average. But even after such an average $Q$ is still highly sensitive to geometric details of the system. The statistical approach consists of considering an ensemble of $Q$ matrices, corresponding to a whole class of similar systems. Universal results can then be obtained. This is the random matrix theory (RMT) approach \cite{sokolov,savin,savin2,fyodorov}. 

For concreteness, consider a limited two-dimensional region where the dynamics is chaotic, coupled with the external world by means of a small opening, modelled as an infinite waveguide of constant cross section. At any given energy, only a finite number of quantum transversal stationary states, $M$, can be populated, called open channels. 

Besides $M$, the only other parameter of the problem is the classical dwell time, $\tau_D$, which measures how much time, on average, a particle spends in the chaotic region, at the given energy and with the given opening. More specifically, if the classical phase-space is filled with initial conditions, the total mass remaining in the region after time $t$ obeys an exponential decay, $e^{-t/\tau_D}$.

The last ingredient of the theory is the information of whether the dynamics is or is not invariant under time-reversal and spin-rotation. A charged particle in a magnetic field would be the prime example of broken time-reversal symmetry (TRS). It is traditional to identify these three cases by means of the Dyson parameter $\beta$, with $\beta=2$ corresponding to systems without TRS, and $\beta=1$ or $\beta=4$ corresponding to TRS with presence or absence of spin-rotation symmetry, respectively. We shall also use the Jack parameter $\alpha$, which is simply
\be \alpha=\frac{2}{\beta},\ee
and the related parameter
\be b=\alpha-1.\ee

\section{Statistical properties} 

When $Q$ is regarged as a random matrix, its eigenvalues are a set of correlated random variables. These eigenvalues, known as the proper time delays, contain information about the duration of the scattering process, and the challenge is to extract statistical information from them. Most of the attention has been focused, over the years, on its normalized trace, the Wigner time delay, 
\be \tau_W=\frac{1}{M}\Tr(Q). \ee
Its average $\langle \tau_W\rangle=\frac{2\pi\hbar\overline{\rho}}{M}=\frac{\tau_H}{M}=\tau_D$
is related to the mean density of states $\overline{\rho}$, the Heisenberg time $\tau_H$ and the dwell time $\tau_D$. The variance and higher moments of $\tau_W$, along with its entire distribution and more refined statistical properties of the matrix $Q$ have been under intense investigation \cite{brouwer1,brouwer2,mezzadri1,mezzadri2,mezzadri3,arguelo,novaes,dist1,dima1}. We refer the reader to the recent review article \cite{texier}. In particular, the tail of the distribution of $\tau_W$ is of the form  \cite{sokolov,savin,savin2,fyodorov}
\be\label{rho} \rho(\tau_W)\sim \tau_W^{-2-\beta M/2}\quad (\tau_W\to\infty).\ee

We should mention that semiclassical approximations to the problem are also possible, in which elements of the $S$ matrix are written in terms of sums over classical scattering trajectories. This has lead \cite{BK1,BK2,novaes2,efficient} to the leading orders in $1/M$ expansions, even in the presence of tunnel barriers \cite{tunnel1,tunnel2}, and recently to the exact result for $\langle \Tr(Q^n)\rangle$ when $\beta=2$ \cite{meunovo}.

When studying spectral statistics, the power traces $\mathcal{T}_k={\rm Tr}(Q^k)$ and the more general power sum symmetric polynomials
\be p_\mu(Q)=\prod_{i=1}^{\ell(\mu)}\mathcal{T}_{\mu_i},\ee
are particularly important figures of merit, as they contain information about variance and higher moments of the Wigner time delay and other moments of the eiganvalue density of the matrix $Q$. Here we denote by $\mu=(\mu_1,\mu_2,...,\mu_{\ell(\mu)})$ an integer partition, i.e. a non-decreasing sequence of $\ell(\mu)$ positive integers.

For broken time-reversal symmetry, a concise and explicit formula is available for $\langle p_\mu(Q) \rangle$ that works for general $\mu$ and $M$ \cite{novaes}. Such result is currently missing for systems with intact TRS. In the present work, we use the theory of Jack polynomials to solve this problem. In fact, our final expressions hold for general $\beta$ (or, equivalently, $\alpha$). That is our main result, presented below in Eqs.(\ref{main})-(\ref{main2}). 

As applications, we derive from that result two interesting conjectures, which generalize previous results about joint cumulants and about the tails of probability distributions.

\section{Conjecture on joint cumulants} 

Joint cumulants are an important way to characterize the statistical properties of a set of random variables. Given $\ell$ variables, $\{y_{\rho_1},...,y_{\rho_\ell}\}$, not necessarily distinct, their joint cumulant is 
\be\label{cumu} K(\rho_1,...,\rho_\ell)=\sum_\pi (|\pi|-1)!(-1)^{|\pi|-1}\prod_{b \in \pi}\left\langle \prod_{i \in b}y_i\right\rangle,\ee 
where the sum over $\pi$ runs through all partitions of the set $\{\rho_1,...,\rho_\ell\}$, $b$ runs through the list of all blocks of the partition $\pi$, and $|\pi|$ is the number of parts in the partition. 

We choose $y_k=p_{(k)}(Q)=\mathcal{T}_k$, so the covariances will be given in terms of power sums. For example, $K(\rho_1,\rho_2)$ is the covariance $\langle p_{(\rho_1,\rho_2)}\rangle-\langle p_{(\rho_1)}\rangle\langle p_{(\rho_2)}\rangle$, while
\begin{multline} K(3,2,1)=\langle p_{(3,2,1)}\rangle -\langle p_{(3,2)}\rangle\langle p_{(1)}\rangle-\langle p_{(3,1)}\rangle\langle p_{(2)}\rangle\\
-\langle p_{(2,1)}\rangle\langle p_{(3)}\rangle+2\langle p_{(3)}\rangle\langle p_{(2)}\rangle\langle p_{(1)}\rangle,\end{multline}
and
\be K(1,1,1)=\langle p_{(1,1,1)}\rangle -3\langle p_{(1,1)}\rangle\langle p_{(1)}\rangle+2\langle p_{(1)}\rangle^3.\ee

In \cite{CMSV} it was conjectured that in the large-$M$ expansion of $K(\rho_1,...,\rho_\ell)$, all terms have positive integer coefficients, when $\alpha\in\{1,2\}$ (this had previously been observed for the leading order term \cite{cunden,leading}). This conjecture was later proved for systems with broken TRS in \cite{proofs1,proofs2}. Based on our main result, we hereby propose an even stronger conjecture: \emph{the coefficients in the large-$M$ expansion of $K(\rho_1,...,\rho_\ell)$ are polynomial in the variable $(\alpha-1)$ with nonnegative integer coefficients}.

More concretely, with $b=\alpha-1$ and $|\rho|=n$, if we denote 
\be\label{conjec}  \frac{M^{\ell-2}}{\tau_D^n}K(\rho_1,...,\rho_\ell)=\alpha^{\ell-1}\sum_{d=0}^\infty \frac{F_{\rho,d}(b)}{M^d},\ee
then we conjecture that $F_{\rho,d}(b)$ is a polynomial of degree $d$ with nonnegative integer coefficients (see below for examples).

This conjecture, if true, obviously implies the result that is known to be valid for $b=0$ \cite{CMSV,cunden,leading,proofs1,proofs2}. Perhaps these polynomials $F_{\mu,d}(b)$ may have some interpretation in terms of the enumeration of non-orientable branched coverings of the sphere studied in \cite{branch1,branch2,branch3}. In fact, this connection has already been stablished for the $b=0$ case in \cite{grava1}.

When (\ref{cumu}) is inverted and $\langle p_\mu(Q)\rangle$ is written in terms of joint cumulants, the coefficients are all positive integers. Therefore, our conjecture implies an analogous one for power sums (see Eq.(\ref{conjp}) below). 

\section{Conjecture on tails of distributions} 

Having exact results for $\langle (\mathcal{T}_k)^n\rangle$, we can investigate its largest pole as a function of $M$, i.e. how large does $M$ have to be to guarantee existence and analyticity of this quantity. This in turn gives an estimate about the tail of the corresponding probability distribution, because if $\rho(\mathcal{T}_k)\sim \mathcal{T}_k^{-f_k}$, then $\langle (\mathcal{T}_k)^n\rangle$ is well defined only if $n<f_k-1$.

The explicit formulas, rational functions of $M$, are too long to display, but if we define a matrix $L$ such that $L_{k,n}$ is this largest pole, then what we find for the smallest values of $k$ and $n$ is
\be \frac{L}{\alpha}=\left( \begin{array}{ccccc}
0 & 1 & 2 & 3&\\
1 & 3 & 5 & 7&\\
2 & 5 & 8 & & \cdots \\ 
3 & 7 &  & & \\
4 & 9 &  & & \\
  & \vdots & & & \ddots 
\end{array} \right).\ee
These observations support the following conjecture: \emph{The largest pole of $\langle (\mathcal{T}_k)^n\rangle$ as a function of $M$ is given by $L_{k,n}=\alpha(kn-1)$. Hence, the tail of the corresponding probability distribution is given by}
\be \rho(\mathcal{T}_k)\sim \mathcal{T}_k^{(-1-k-M/\alpha)/k}\quad (\mathcal{T}_k\to\infty).\ee

The above estimate should be possible to test numerically and maybe proved using Coulomb gas methods \cite{dist1}. It generalizes Eq.(\ref{rho}), which corresponds to the simplest case $k=1$ and has already been proved.

\section{Jack polynomials} 

In order to perform our calculations, we shall, instead of dealing with power sums, employ a different family of symmetric polynomials, the Jack polynomials $P_\lambda^{(\alpha)}(X)$,depending on variables $X=\{x_1,...,x_M\}$ and indexed by the parameter $\alpha$ and by an integer partition $\lambda$. As usual, we denote by $|\lambda|=\sum_i\lambda_i$ and write $\lambda\vdash n$ when $|\lambda|=n$.  

We choose the monic normalization for $P_\lambda^{(\alpha)}$, such that the coefficient of the monomial symmetric polynomial $m_\lambda$ is unit. These polynomials have several remarkable properties (we refer the reader to the standard reference \cite{macdonald}) and have found numerous applications, e.g. \cite{vinet,forr,app1,app2}. Among other things, the set $\{P_\lambda^{(\alpha)},\lambda\vdash n\}$ is a basis for the space of degree-$n$ homogeneous symmetric polynomials. In particular, the coefficients in the expansion
\be p_\mu(Q)=\sum_{\lambda\vdash n}\chi^{(\alpha)}_{\mu\lambda} P_\lambda^{(\alpha)}(Q)\ee
are known as the Jack characters \cite{charac1,charac2,charac3,charac4} (in the special case $\alpha=1$ the Jack polynomials become Schur polynomials, and the corresponding characters become the irreducible characters of the permutation group). For example,
\be p_{(2)}(Q)=\frac{1}{\alpha+1}(P_{(2)}^{(\alpha)}(Q)-P_{(1,1)}^{(\alpha)}(Q)),\ee
and
\be p_{(1,1)}(Q)=\frac{1}{\alpha+1}(P_{(2)}^{(\alpha)}(Q)+\alpha P_{(1,1)}^{(\alpha)}(Q)).\ee

Jack polynomials are convenient in the present context because of an important result known as the Selberg-Jack integral \cite{kaneko,kadell,selberg}, which says that 
\be \int_0^\infty e^{-z\Tr(X)}\det(X)^{a-1}|\Delta(X)|^{2/\alpha} P_\lambda^{(\alpha)}(X)
dX\ee
is equal to 
\be\label{selberg}\frac{P_\lambda^{(\alpha)}(1^M)}{z^{|\lambda|+Ma+\frac{M(M-1)}{\alpha}}}\prod_{j=1}^M\frac{\Gamma(\lambda_j+a+\frac{(M-j)}{\alpha})\Gamma(1+\frac{j}{\alpha})}{\Gamma(1+\frac{1}{\alpha})},\ee
where $\Gamma$ is the Gamma function, $1^M$ denotes the identity matrix in dimension $M$ and the so-called Vandermonde determinant is given by
\be \Delta(X)=\prod_{1\le i<j\le M}(x_j-x_i).\ee This is a deep generalization of the Euler beta integral. 

The value of the Jack polynomial at the identity is well known to be \cite{macdonald}
\be P_\lambda^{(\alpha)}(1^M)=\frac{\alpha^{|\lambda|}}{H_\lambda^{(\alpha)}}\prod_{j=1}^{\ell(\lambda)}\frac{\Gamma(\lambda_j+\frac{(M-j+1)}{\alpha})}{\Gamma(\frac{(M-j+1)}{\alpha})},\ee
where
\be H_\lambda^{(\alpha)}=\prod_{i=1}^{\ell(\lambda)}\prod_{j=1}^{\lambda_i}(\alpha(\lambda_i-j)-i+1+\lambda'_j),\ee
with $\lambda'$ denoting the partition conjugate to $\lambda$, obtained by 
transposing the Young diagram representation of the partition, i.e. exchanging its rows and columns (for example, $(3,2)'=(2,2,1)$).

\section{Calculations} 

The joint probability distribution of the eigenvalues of $Q$ is known. When the coupling between the chaotic region and the lead is perfectly transparent (see \cite{tunnel} for a treatment taking into account imperfect coupling) it is given in terms of $\gamma=Q^{-1}$, the inverse of the time-delay matrix, as \cite{brouwer2}
\be \frac{1}{\mathcal{Z}}e^{-\frac{\beta}{2} M\tau_D\Tr(\gamma)}\det(\gamma)^{\beta M/2}|\Delta(\gamma)|^\beta,\ee
depending on the Dyson parameter and the dwell time, with normalization
\be \mathcal{Z}=\int_0^\infty e^{-\frac{\beta}{2} M\tau_D\Tr(\gamma)}\det(\gamma)^{\beta M/2}|\Delta(\gamma)|^\beta d\gamma.\ee As is typical of random matrix ensembles, the correlation between the eigenvalues of $\gamma$ arises precisely through the presence of the Vandermonde $\Delta(\gamma)$.

Taking $\lambda$ to be the empty partition and choosing $a$ and $z$ appropriately, we can use (\ref{selberg}) to find for the normalization factor $\mathcal{Z}$ the value
\be \left(\frac{\alpha}{M\tau_D}\right)^{\frac{M}{\alpha}(2M+\alpha-1)}\prod_{j=1}^M\frac{\Gamma(1+\frac{(2M-j)}{\alpha})\Gamma(1+\frac{j}{\alpha})}{\Gamma(1+\frac{1}{\alpha})}.\ee

The average value of a Jack polynomial in the eigenvalues of $Q$ is given by
\be \frac{1}{\mathcal{Z}}\int e^{-\frac{M}{\alpha}\tau_D\Tr(\gamma)}\det(\gamma)^{M/\alpha}|\Delta(\gamma)|^{2/\alpha} P_\lambda^{(\alpha)}(\gamma^{-1})d\gamma.\ee The Selberg-Jack integral cannot be applied directly here. However, it is possible to show that 
\be \label{relation}\det(\gamma)^{M/\alpha} P_\lambda^{(\alpha)}(\gamma^{-1})=P_{\widetilde{\lambda}}^{(\alpha)}(\gamma),\ee
where 
\be \widetilde{\lambda}=\left(\frac{M}{\alpha}-\lambda_M,\frac{M}{\alpha}-\lambda_{M-1},...,\frac{M}{\alpha}-\lambda_1\right).\ee
This list is not necessarily an integer partition, but this presents no difficulty since (\ref{selberg}) depends on the parts of $\lambda$ in an analytic way. Notice that 
$|\widetilde{\lambda}| =\frac{M^2}{\alpha}-n$. 

The average we want thus becomes 
\be \langle  P_\lambda^{(\alpha)}(Q)\rangle=\frac{1}{\mathcal{Z}}\int e^{-\frac{M}{\alpha}\tau_D\Tr(\gamma)}|\Delta(\gamma)|^{2/\alpha} P_{\widetilde{\lambda}}^{(\alpha)}(\gamma)d\gamma.\ee
According to (\ref{selberg}), this is equal to
\be\label{res1} \left(\frac{M\tau_D}{\alpha}\right)^{n}P_{\widetilde{\lambda}}^{(\alpha)}(1^M)\prod_{j=1}^M\frac{\Gamma(-\lambda_{N-j+1}+1+\frac{(2M-j)}{\alpha})}{\Gamma(1+\frac{(2M-j)}{\alpha})}.\ee Moreover, taking $\gamma=1^M$ in Eq.(\ref{relation}) we get immediately
\be\label{equal} P_\lambda^{(\alpha)}(1^M)=P_{\widetilde{\lambda}}^{(\alpha)}(1^M).\ee

Let us introduce $\alpha$-dependent generalized rising and falling factorials as
\be [M]^\lambda_{(\alpha)}=\alpha^{|\lambda|}\prod_{j=1}^{\ell(\lambda)}\frac{\Gamma(\lambda_j+\frac{(M-j+1)}{\alpha})}{\Gamma(\frac{(M-j+1)}{\alpha})},\ee and
\be [M]_\lambda^{(\alpha)}=\alpha^{|\lambda|}\prod_{j=1}^{\ell(\lambda)}\frac{\Gamma(\frac{(M+j)}{\alpha})}{\Gamma(-\lambda_j+\frac{(M+j)}{\alpha})}.\ee If we exchange the dummy variable $j$ in (\ref{res1}) for $M-j+1$, and make use of Eq.(\ref{equal}), we find 
\be\label{main} \langle  P_\lambda^{(\alpha)}(Q)\rangle=(M\tau_D)^{n}\frac{1}{H_\lambda^{(\alpha)}}\frac{[M]^\lambda_{(\alpha)}}{[M-1+\alpha]_\lambda^{(\alpha)}}.\ee
Therefore, average values of Jack polynomials are very simple, e.g.
\begin{align} 
\langle  P_{(3)}^{(\alpha)}(Q)\rangle&=\frac{(M\tau_D)^3}{(2\alpha+1)(\alpha+1)}\frac{(M+\alpha)(M+2\alpha)}{(M-\alpha)(M-2\alpha)},\\
\langle  P_{(2,1)}^{(\alpha)}(Q)\rangle&=\frac{(M\tau_D)^3}{\alpha+2}\frac{(M+\alpha)(M-1)}{(M-\alpha)(M-1)},\\
\langle  P_{(1,1,1)}^{(\alpha)}(Q)\rangle&=\frac{(M\tau_D)^3}{6}\frac{(M-1)(M-2)}{(M+1)(M+2)}.\end{align}

The more familiar power sum symmetric functions then become
\be\label{main2} \langle p_\mu(Q)\rangle=(M\tau_D)^{n}\sum_{\lambda\vdash n}\frac{\chi^{(\alpha)}_{\mu\lambda}}{H_\lambda^{(\alpha)}}\frac{[M]^\lambda_{(\alpha)}}{[M-1+\alpha]_\lambda^{(\alpha)}}. \ee These quantities are in fact not of order $M^n$ for large $M$, because of cancellations (Jack characters may be negative). In fact, it seems $\langle p_\mu(Q)\rangle$ is of order $M^{\ell(\mu)}$, so the normalized quantity 
$\tau_D^{-n}M^{-\ell(\mu)}\langle p_\mu(Q)\rangle$ tends to a constant for large $M$.

\section{Examples} 

Not much is known about the Jack characters $\chi^{(\alpha)}_{\mu\lambda}$, but some packages are available that allow their computation for small $n$ (see \cite{maple1,maple3}). Using a computer, we obtain, for example,
\begin{align} \frac{\langle p_{(3)}(Q)\rangle}{\tau_D^3M}&=\frac{6M^4}{(M-2\alpha)(M-\alpha)(M+1)(M+2)},\\
\frac{\langle p_{(2,1)}(Q)\rangle}{\tau_D^3M^2}&=\frac{2M^2[M(M+2)-2(M-1)\alpha]}{(M-2\alpha)(M-\alpha)(M+1)(M+2)},\\
\frac{\langle p_{(1,1,1)}(Q)\rangle}{\tau_D^3M^3}&=\frac{[A\alpha^2-B\alpha+C]}{(M-2\alpha)(M-\alpha)(M+1)(M+2)},\end{align}
where $A=2(M-1)(M-2)$, $B=3M(M-1)(M+2)$, and $C=M^2(M+1)(M+2)$.

Curiously, the time delay moments may develop singularities for fractial $\alpha$ if the number of channels is small. For instance, if $M=3$ we have 
\be \frac{\langle p_{(2,1)}(Q)\rangle}{\tau_D^3}=\frac{81}{10}\frac{15-4\alpha}{(3-\alpha)(3-2\alpha)}.\ee This is finite at $\alpha=1$ and $\alpha=2$, but undefined at $\alpha=3/2$. Whether this corresponds to any actual physics is not clear and deserves further investigation. On the other hand, $\langle p_{\mu}(Q)\rangle$ is always finite if $M>\alpha|\mu|$.

Turning to the joint cumulants, the simplest ones are
\begin{align} K(1,1)&=\frac{2M^2\alpha}{(M-\alpha)(M+1)},\\
K(2,1)&=\frac{12M^4\alpha}{(M-\alpha)(M-2\alpha)(M+1)(M+2)},\\
K(1,1,1)&=\frac{24M^3\alpha^2}{(M-\alpha)(M-2\alpha)(M+1)(M+2)}.\end{align}
The simplicity of these expressions is misleading; when $\ell>3$ or $|\rho|>3$ they become rather cumbersome. 

As already mentioned, it has been conjectured that the large-$M$ expansion of $\langle p_\mu(Q)\rangle$ should have positive integer coefficients for $\alpha\in\{1,2\}$. Based on our calculations, we put forth a more general conjecture: if we denote 
\be\label{conjp}  \frac{1}{\tau_D^nM^{\ell(\mu)}}\langle p_{\mu}(Q)\rangle=\sum_{d=0}^\infty \frac{G_{\mu,d}(b)}{M^d},\ee
then $G_{\mu,d}(b)$ is a polynomial of degree $d$ with nonnegative integer coefficients.

Let us present some examples of these polynomials. For $\mu=(3)$ we have $G_{(3),0}=6$, $G_{(3),1}=18b,$ and
\be G_{(3),2}=42b^2+30b+30.\ee 
For $\mu=(2,1)$ we have $G_{(2,1),0}=2$, $G_{(2,1),1}=2b$, and
\be  G_{(2,1),2}=2b^2+14b+14.\ee
For $\mu=(1,1,1)$ we have $G_{(1,1,1),0}=1$, $G_{(1,1,1),1}=6b+6$, and
\be  G_{(1,1,1),2}=6b^2+6b.\ee

On the other hand, here are some examples of the polynomials defined in Eq.(\ref{conjec}) for the asymptotics of the joint cumulants. For $\rho=(1,1)$, i.e. the covariance of the Wigner time delay $\tau_W$, we have $F_{(1,1),0}=2$, $F_{(1,1),1}=2b$, and
\be F_{(1,1),2}=2b^2+2b+2.\ee 
For $\rho=(2,1)$, we have $F_{(2,1),0}=12$, $F_{(2,1),1}=36b$, and
\be  F_{(2,1),2}=84b^2+60b+60.\ee
For $\rho=(1,1,1)$, we have  $F_{(1,1,1),0}=24$, $F_{(1,1,1),1}=72b$, and
\be  F_{(1,1,1),2}=168b^2+120b+120.\ee

We have also noted that both families of polynomials, $F_{\rho,d}$ and $G_{\mu,d}$, have an interesting symmetry property, when written in terms of $\alpha$: they are self-reciprocal, meaning they are invariant, up to an overall power of $\alpha$, under the replacement $\alpha\mapsto \alpha^{-1}$. In terms of the Dyson parameter, this symmetry becomes $\beta\mapsto \frac{4}{\beta}$, interchanging $\beta=1$ and $\beta=4$ while leaving $\beta=2$ invariant.

\section{Conclusions} 

Using the joint probability distribution of inverse delay times and the theory of Jack polynomials, we computed all spectral statistics of the Wigner time delay matrix $Q$, for arbitrary channels number $M$ and for all values of the Dyson paramter $\beta$. We have found that, for small values of $M$, the observables develop singularities at fractional values of $\beta$, but possible physical consequences of this are yet to be investigated.

We have also presented two conjectures. One is related to the large-$M$ expansion of joint cumulants of traces of powers of $Q$, which generalizes and implies a previous conjecture of Cunden, Mezzadri, Vivo and Simm. The other is related to the tail of the probability distribution of ${\rm Tr}(Q^k)$, and generalizes to general $k$ the known result for $k=1$. Perhaps this can be proved using Coulomb gas methods such as in \cite{dist1}.

That some results involving Jack polynomials have nice positivity and integrality properties when expressed in terms of the modified parameter $b=\alpha-1$ is a phenomenon that has been noticed before, in connection with a conjecture of Goulden and Jackson \cite{GJ} related to the enumeration of non-orientable branched coverings of the sphere \cite{branch1,branch2,branch3}. Perhaps our polynomials $F_{\mu,d}(b)$ and $G_{\rho,d}(b)$ may have some interpretation in terms of such enumeration. In fact, this connection has already been stablished for the $\alpha=1$ case in \cite{grava1}. This topic deserves further study.

\acknowledgments

Financial support from CNPq, grant 306765/2018-7, is gratefully acknowledged. I would like to thank John Stembridge and Doron Zeilberger for making their symmetric polynomial codes freely available. The possibility of self-reciprocity for the polynomials $F_{\rho,d}$ and $G_{\mu,d}$ was suggested to us by Fabio Cunden. Computer codes for the calculations presented here are available upon request.

\end{document}